\documentclass[prd,amssymb,eqsecnum,showpacs,preprint]{revtex4}
\usepackage{epsfig}
\newcommand{\beq}{\begin{equation}}
\newcommand{\eeq}{\end{equation}}
\newcommand{\bqa}{\begin{eqnarray}}
\newcommand{\eqa}{\end{eqnarray}}
\newcommand{\fr}{\frac}
\newcommand{\nb}{\nonumber}

\begin{document}
\title{Relativistic shells: Dynamics, horizons, and shell crossing}
\thanks{Journal reference: Phys. Rev. D {\bf 66}, 084021 (2002)}
\author{S\'{e}rgio M. C. V. Gon\c{c}alves}
\affiliation{Department of Physics, Yale University, New Haven, Connecticut 06511}
\date{March 10, 2002}
\begin{abstract}
We consider the dynamics of timelike spherical thin matter shells in vacuum. A general formalism for thin shells matching two arbitrary spherical spacetimes is derived, and subsequently specialized to the vacuum case. We first examine the relative motion of two dust shells by focusing on the dynamics of the exterior shell, whereby the problem is reduced to that of a single shell with different active Schwarzschild masses on each side. We then examine the dynamics of shells with non-vanishing tangential pressure $p$, and show that there are no stable---stationary, or otherwise---solutions for configurations with a strictly linear barotropic equation of state, $p=\alpha\sigma$, where $\sigma$ is the proper surface energy density and $\alpha\in(-1,1)$. For {\em arbitrary} equations of state, we show that, provided the weak energy condition holds, the strong energy condition is necessary and sufficient for stability. We examine in detail the formation of trapped surfaces, and show explicitly that a thin boundary layer causes the apparent horizon to evolve discontinuously. Finally, we derive an analytical (necessary and sufficient) condition for neighboring shells to cross, and compare the discrete shell model with the well-known continuous Lema\^{\i}tre-Tolman-Bondi dust case.
\end{abstract}
\maketitle

\section{Introduction}

The matching of two arbitrary spacetimes along a given hypersurface plays an important role in general relativity, with a rich plethora of applications, such as the dynamics of thin matter shells~\cite{israel67}, construction of cosmological models~\cite{krasinsky97}, collapse of bounded bodies~\cite{mtw}, and wormholes~\cite{visser89}. The standard techniques to achieve such matching are not circumscribed to four-dimensional manifolds, and can be easily applied to higher-dimensional cases, such as $n$-branes in the so-called brane world cosmology~\cite{branecosm}.

Within the context of classical general relativity, the thin shell matching problem was first studied by Sen~\cite{sen24}, Lanczos~\cite{lanczos24} and Darmois~\cite{darmois27}, and later further developed by Israel~\cite{israel66-67}, who produced a coordinate invariant formalism by applying the Gauss-Codacci equations to a non-null three-dimensional hypersurface imbedded in a four-dimensional spacetime. The null case was studied by Barrab\`{e}s and Israel~\cite{barrabes&israel91}. On the shell, Einstein's equations reduce to the Lanczos equation, where the jump of the extrinsic curvature across the shell plays the role of the four-dimensional Ricci tensor, being thus related to the surface stress-energy tensor. Comprehensive reviews of the matching problem in general relativity may be found in~\cite{lichnerowicz95,lake&musgrave96}.

In this paper, we first consider two shells in vacuum, and analyze the dynamics from the viewpoint of the exterior shell, thereby reducing the problem to that of a single shell immersed in two adjacent Schwarzschild spacetimes with different mass parameters on each side. For shells with pressure, we show that (i) a strictly linear barotropic equation of state is incompatible with stable (oscillatory or stationary) solutions, and (ii) for matter with a real local sound speed ($C_{\rm s}\equiv\sqrt{dp/d\sigma}\geq0$) obeying the weak energy condition, the strong energy condition is necessary for stability, regardless of the equation of state. We then study the formation of trapped surfaces in the spacetime, and show explicitly that the existence of a thin matter shell introduces a discontinuity in the apparent horizon curve. The related problem of shell crossings is examined, and an analytical necessary and sufficient condition for neighboring shells (defined in a precise manner) to cross is derived. We further show that, although shell crossing singularities occur in a multi-dust-shell case just as they do in the continuous Lema\^{\i}tre-Tolman-Bondi (LTB) dust case, the former cannot be taken as the discrete analogue of the latter, since, e.g., the energy density remains finite at discrete shell crossings, whereas it diverges (together with some curvature components) in the LTB case. Physically, this is related to the fact that individual shells move geodesically in the LTB spacetime (i.e., proper time equals comoving time, for each shell), but the same does fails to hold in the discrete case: dust particles do move geodesically on each shell, but, from the viewpoint of the four-dimensional spacetime, they are accelerated.

The paper is organized as follows: Section II derives a general formalism for the matching of two arbitrary spherical spacetimes across a timelike thin shell with arbitrary matter content. In Sec. III we specialize to the case of two dust shells in vacuum; the general relativistic equation of motion for the exterior shell is derived, as well as its Newtonian analogue. The well-known case of a single dust shell in vacuum is trivially recovered. In Sec. IV, the dynamics of a dust shell with a LTB interior is discussed, and the boundary layer case of closed Lema\^{\i}tre-Friedmann-Robertson-Walker (LFRW) models is readily obtained as special case. Section V studies shells with a non-vanishing tangential pressure component, and several stability results are produced. In Sec. VI, we derive the condition for formation of apparent horizons in an arbitrary spherical spacetime, and then examine the particular case of a dust shell collapsing in Schwarzschild background. Section VII discusses the occurrence of shell crossing in a spacetime with a finite number of shells, derives an explicit condition for shell crossing of neighboring shells, and compares it with the continuous dust LTB case. Section VIII concludes with a summary and discussion.

Natural geometrized units, in which $G=c=1$, are used throughout.

\section{Matching formalism for arbitrary spherical spacetimes}

The complete four-dimensional spacetime consists of an interior region ${\mathcal M}_{-}$ connected to an exterior ${\mathcal M}_{+}$ by a timelike three-dimensional thin shell $\Sigma$. The $\mathcal{M}_{\pm}$ regions are characterized by the spherical metric:
\beq
ds^{2}_{\pm}=-h_{\pm}^{2}dt_{\pm}^{2}+f_{\pm}^{2}dr^{2}+r^{2}d\Omega^{2}, \label{metric}
\eeq
where $h_{\pm}, f_{\pm}$ are functions of $t_{\pm}$ and $r$, and the coordinate systems $\{x_{\pm}^{\mu}\}$ are adopted. On $\Sigma$ there is a natural holonomic basis $\{e_{(a)}\}$ given by
\beq
e^{\mu}_{(a)}|_{\pm}=\fr{\partial x_{\pm}^{\mu}}{\partial \xi^{a}},
\eeq
where $\{\xi^{a}, a=0,1,2\}$ are intrinsic coordinates on $\Sigma$. The induced three-metric $\gamma_{ab}$ on $\Sigma$ is then
\beq
\gamma_{ab}=g_{\mu\nu}e^{\mu}_{(a)}e^{\nu}_{(b)},
\eeq
and it is the same on both sides of $\Sigma$, since the four-metric must be continuous across it. The surface $\Sigma$ is parametrically defined by
\beq
\Phi(x_{\pm}^{\mu})=r-r_{\Sigma}(t_{\pm})=0.
\eeq
Continuity of $g_{\mu\nu}$ across $\Sigma$ requires $r_{\Sigma}(t_{\pm})=R(\tau)$, whence
\bqa
ds^{2}_{\Sigma}&=&-d\tau^{2}+R^{2}(\tau)d\Omega^{2}, \\
\fr{dt_{\pm}}{d\tau}&=&h^{-1}_{\pm}\sqrt{1+f^{2}_{\pm}\dot{R}^{2}}\equiv\eta_{\pm}, \label{eta}
\eqa
where $\dot{R}\equiv dR/d\tau$, and $\tau$ is the proper time measured by a comoving observer with the shell, with four-velocity
\beq
u_{\pm}^{\mu}=\eta_{\pm}\delta^{\mu}_{t_{\pm}}+\dot{R}\delta^{\mu}_{r}.
\eeq
The spacelike (outward pointing) unit normal to $\Sigma$ is
\beq
n^{\mu}_{\pm}=g^{\mu\nu}_{\pm}\alpha^{-1}\partial^{\pm}_{\nu}\Phi=h_{\pm}^{-1}f_{\pm}\dot{R}\delta^{\mu}_{t_{\pm}}+f^{-1}_{\pm}h_{\pm}\eta_{\pm}\delta^{\mu}_{r},
\eeq
where $\alpha^{2}\equiv n_{\pm}^{\mu}n^{\pm}_{\mu}$, and $n^{\pm}_{\mu}u^{\mu}_{\pm}=0$. We shall henceforth drop the $\pm$ subscript for clarity and without detriment, since all the formulae will apply equally to both sides of $\Sigma$.

The normal extrinsic curvature, $K_{ab}$, is~\cite{israel66-67}
\beq
K_{ab}:=-n_{\mu}e^{\nu}_{(b)}\nabla_{\nu}e^{\mu}_{(a)}=-n_{\sigma}\left(\fr{\partial^{2}x^{\sigma}}{\partial\xi^{a}\partial\xi^{b}}+\Gamma^{\sigma}_{\mu\nu}\fr{\partial x^{\mu}}{\partial\xi^{a}}\fr{\partial x^{\nu}}{\partial\xi^{b}}\right),
\eeq
which is related to the (yet unspecified) surface stress-energy tensor $S_{ab}$ on $\Sigma$ via the Lanczos equation~\cite{israel66-67}
\beq
[K_{ab}]-\gamma_{ab}[K]=-8\pi{S_{ab}} \;\; \Leftrightarrow\;\; [K_{ab}]=-8\pi(S_{ab}-\fr{1}{2}S\gamma_{ab}), \label{lzc}
\eeq
where
\beq
[K_{ab}]\equiv K_{ab}^{+}-K_{ab}^{-}, \;\;\; [K]\equiv\gamma^{ab}[K_{ab}], \;\;\; S\equiv\gamma_{ab}S^{ab}.
\eeq
The four-dimensional stress-energy tensor associated with $\Sigma$ can be written as a distribution as
\beq
T_{\Sigma}^{\,u\nu}=S^{ab}e^{\mu}_{(a)}e^{\nu}_{(b)}|\alpha|\delta(\Phi).
\eeq
The Lanczos equation (for the shell), together with the (exterior) Einstein equations, lead to the standard Israel junction conditions~\cite{israel66-67}:
\bqa
S^{ab}\{K_{ab}\}&\equiv&\fr{1}{2}(K_{ab}^{+}+K_{ab}^{-})S^{ab}=[T_{\mu\nu}n^{\mu}n^{\nu}], \label{admc} \\
^{(3)}\nabla_{b}S^{b}_{a}&=&[e^{\mu}_{(a)}T_{\mu\nu}n^{\nu}]. \label{hamc}
\eqa
These two equations are identities that must be satisfied throughout the time development of the shell, and not genuine dynamical equations, since they follow from the momentum and Hamiltonian constraints imposed on $\Sigma$~\cite{israel66-67}; the dynamics is encoded in the Lanczos equation (\ref{lzc}).

With the metric (\ref{metric}), the non-vanishing components of $K_{ab}$ are:
\bqa
K_{\tau\tau}&=&-\fr{f}{h\eta}\left(\ddot{R}+\dot{R}\fr{\dot{f}}{f}\right)-\fr{1}{f}D_{\perp}f+\fr{\eta}{f}\left(\fr{f'}{fh\eta^{2}}-h'\right), \label{ktt} \\
K_{\theta\theta}&=&\sin^{-2}\theta K_{\phi\phi}=\fr{Rh\eta}{f}, \label{kthth}
\eqa
where $D_{\perp}\equiv n^{\mu}\nabla_{\mu}$ is the normal derivative in the $n^{\mu}$ direction, and $'\equiv\partial_{r}$.

\section{Two collapsing Schwarzschild shells}

We shall examine here the case of two neighboring thin shells of dust. Specifically, the model consists of an inner shell with gravitational mass $m_{-}$ and an outer shell with $m_{+}$, in an otherwise empty spacetime. An immediate consequence of Birkhoff's theorem is that the two-shell case can be reduced to that of a single shell with appropriate adjoint spacetime metrics. If one is interested in the dynamics of the inner shell, the problem is that of a single shell with a Minkowski interior (to ensure regularity of the metric at the center) and a Schwarzschild exterior with mass $M=m_{-}+m_{+}$. Focusing on the exterior shell, the problem reduces to that of a single shell with an interior Schwarzschild metric with mass $m_{-}$ and Schwarzschild exterior with mass $M$. We shall henceforth adopt this latter viewpoint. The metrics in $\mathcal{M}_{\pm}$ are then:
\beq
ds^{2}_{\pm}=-h_{\pm}^{2}dt^{2}+h^{-2}_{\pm}dr^{2}+r^{2}d\Omega^{2},
\eeq
with
\beq
h_{-}=\sqrt{1-\fr{2m_{-}}{r}}, \;\;\; h_{+}=\sqrt{1-\fr{2M}{r}}, \;\;\; M\equiv m_{-}+m_{+}. \label{schwp}
\eeq
In this case, Eqs. (\ref{ktt})-(\ref{kthth}) simplify to
\beq
K_{\tau\tau}=-\fr{\ddot{R}+h'h}{\eta h^{2}}, \;\;\;\; K_{\theta\theta}=Rh^{2}\eta. \label{kschw}
\eeq
Now, let $\sigma$ be the total rest mass of the shell per unit proper area. The stress-energy tensor $S_{ab}$ on $\Sigma$ is then
\beq
S_{ab}=\sigma u_{a}u_{b}=\sigma e^{\mu}_{(a)}e^{\nu}_{(b)}u_{\mu}u_{\nu}=\sigma\delta^{\tau}_{a}\delta^{\tau}_{b},
\eeq
and the total rest mass of the shell is
\beq
m_{\Sigma}=\int_{\Sigma} S_{ab}u^{a}u^{b}\sqrt{\mbox{det}\gamma}{\rm d}\theta \wedge {\rm d}\phi=4\pi R^{2}\sigma. \label{rmass}
\eeq
The conservation equation (\ref{hamc}) gives
\beq
\fr{\dot{\sigma}}{\sigma}+2\fr{\dot{R}}{R}\equiv\fr{\dot{m}_{\Sigma}}{4\pi}=0,
\eeq
i.e., the proper mass of the shell is conserved during the evolution; microscopically, this simply reflects the fact that the total number of particles in the shell is conserved, and thus the flux three-vector $j^{a}=\sigma u^{a}$ is divergence-free.

The dynamical evolution of the shell can be easily obtained from the $\theta\theta$ component of the Lanczos equation [a straightforward calculation shows that Eq. (\ref{admc}) is equivalent to $[K_{\theta\theta}]=-4\pi\sigma R^{2}$, and the $\tau\tau$ component of the Lanczos equation does not yield additional information---its first integral is automatically satisfied by Eq. (\ref{dyn})]:
\beq
[K_{\theta\theta}]=-4\pi\sigma R^{2}. \label{kth2}
\eeq
>From Eqs. (\ref{schwp}), (\ref{rmass}), and (\ref{kth2}) we obtain
\beq
\dot{R}^{2}=\left(\fr{m_{+}}{m_{\Sigma}}\right)^{2}-1+\fr{M+m_{-}}{R}+\fr{m_{\Sigma}^{2}}{4R^{2}}. \label{dyn}
\eeq
This is the full general relativistic equation governing the motion of the shell $\Sigma$ with active gravitational mass $m_{+}$ in an interior Schwarzschild background with active gravitational mass $m_{-}$. Since both the gravitational and rest masses of the shell remain constant during the evolution, we can introduce the dimensionless constant $k\equiv m_{+}/m_{\Sigma}$, and rewrite the dynamical equation solely in terms of the gravitational masses:
\beq
\dot{R}^{2}=k^{2}-1+\fr{M+m_{-}}{R}+\fr{m_{+}^{2}}{4k^{2}R^{2}}. \label{dyn2}
\eeq
Based on its Newtonian analogue (derived below), one may interpret the terms in this equation as follows: $k^{2}-1$ is the total specific (i.e., per unit gravitational mass $m_{+}$) binding energy of the system, the second term is the potential energy, and the third term is a self-binding energy for $\Sigma$, which is a relativistic correction to the Newtonian case. We note that, since energy (potential, or otherwise) gravitates, the proper mass of the shell will only coincide with its active gravitational mass if the total binding energy of the system vanishes, when $k=1$. For given initial data $\{\tau_{\rm i}; R(\tau_{\rm i}), \dot{R}(\tau_{\rm i})\}$, collapsing or expanding solutions may be obtained, depending on the value of $k\in(0,+\infty)$. For $k\in(0,1)$, potential (negative) energy dominates, and there is a maximum radius at which the shell is momentarily at rest before recollapsing, i.e., the system is gravitationally bound; for $k=1$, the kinetic energy vanishes exactly at spatial infinity, and the system is said to be marginally bound; if $k\in(1,+\infty)$, the kinetic energy is positive-definite at infinity [it ``equals'' $m_{+}(k^{2}-1)/2$] and the system is said to be gravitationally unbound.

\subsection{Newtonian limit}

The Newtonian limit is obtained by the requirement of non-relativistic velocities, $|\dot{R}|\ll1$, which implies~\cite{landau&lifshitz2} the weak-field limit, $\epsilon\equiv m_{\pm}/R<<1$. To first-order in $\epsilon$, Eq. (\ref{dyn2}) reads
\beq
\dot{R}^{2}=k^{2}-1+\fr{M+m_{-}}{R}+{\mathcal O}(\epsilon^{2}). \label{newt}
\eeq
To check that this is indeed the correct limit, we present below a simple Newtonian derivation of the same equation. Let $m_{\mp}$ be the rest masses of the inner and outer shells, respectively. We want to follow the motion of the $m_{+}$ shell in the gravitational potential of the $m_{-}$ shell. By Gauss's theorem, this reduces to the problem of finding the dynamical equation of a spherical shell $\Sigma$ with radius $r=R$ and mass $m_{+}$, subject to the gravitational potential of a point mass $m_{-}$ located at $r=0$. The action for such system is
\beq
S=\int L(R,\dot{R},t) dt=\int \left[T-(U_{-}+U_{\Sigma})\right] dt,
\eeq
where
\bqa
T&=&\fr{1}{2}m_{+}\dot{R}^{2}, \\
U_{-}&=&-\fr{m_{-}m_{+}}{R}, \\
U_{\Sigma}&=&\fr{1}{2}\int_{\Sigma} \Phi_{\Sigma}(x)\rho(x)d^{3}x \nb \\
&=&2\pi\int^{R}_{0}\Phi_{\Sigma}(r)\rho(r)r^{2}dr=-\fr{m_{+}^{2}}{2R},
\eqa
are the kinetic, potential, and self-binding (of the shell) energies, respectively. The function $\Phi_{\Sigma}(r)\equiv-m_{+}/r$ is the self-gravitational potential of the shell, and $\rho(r)\equiv\sigma\delta(r-R)$ is the energy density, where $\sigma=m_{+}/(4\pi R^{2})$ is the shell's surface energy density. Extremizing $\delta S=0$ yields the Euler-Lagrange equation
\beq
\ddot{R}+\fr{2m_{-}+m_{+}}{2R^{2}}=0,
\eeq
which integrates to
\beq
\dot{R}^{2}=\Omega+\fr{M+m_{-}}{R},
\eeq
where $M\equiv m_{+}+m_{-}$, and $\Omega$ is an integration constant, which represents (twice) the total binding energy of the system per unit shell mass. This agrees thus with Eq. (\ref{newt}) upon setting $\Omega=k^{2}-1$.

\subsection{Single dust shell in vacuum}

By setting $m_{-}=0$, we readily obtain the case of a single spherical dust shell in an otherwise vacuum spacetime, originally studied by Israel~\cite{israel67}:
\beq
\dot{R}^{2}=-1+\left(k+\fr{M}{2kR}\right)^{2}=k^{2}-1+\fr{M}{R}+\fr{M^{2}}{4k^{2}R^{2}}. \label{dyn2i}
\eeq
From the first equality above it follows that $2kR(1-k)\leq M$. If $k\geq1$, this is trivially satisfied, but for $k<1$ this imposes an upper bound on $R$:
\beq
R\leq \fr{M}{2k(1-k)}\equiv R_{\rm max}. \label{rmax}
\eeq
As before, if the system is gravitationally bound ($k<1$), any initially expanding shell will reach a maximum radius---uniquely determined by the gravitational mass of the shell and its total binding energy---and then collapse back to $R=0$. Gravitationally unbound shells ($k>1$) will expand to $R\rightarrow\infty$. As noted by Israel, the upper bound (\ref{rmax}) implies that $R_{\rm max}\geq2M$, and hence there are no timelike stationary shells with $R<2M$ (any stationary shell must necessarily be spacelike), as expected. Further details, including exact solutions of Eq. (\ref{dyn2i}), may be found in the original reference~\cite{israel67}.

\section{Dust shell with Lema\^{\i}tre-Tolman-Bondi interior}

Let us now consider the case of a spherical dust shell matched to an interior LTB spacetime~\cite{ltb}, and a Schwarzschild exterior. The interior metric reads
\beq
ds^{2}_{-}=-dt^{2}+F^{2}(t,r)dr^{2}+X^{2}(t,r)d\Omega^{2}, \label{ltbm}
\eeq
where $F\equiv X'/\sqrt{1-w(r)}$, and $4\pi X^{2}$ is the ($t$ dependent) proper area of a spherical shell with coordinate radius $r$. For simplicity, we shall consider a class of LTB metrics given by a separable area radius function, $X(t,r)=a(t)r$. Included in this class are the LFRW cosmological spacetimes, given by $w(r)=Ar^{2}$, where the constant $A$ determines the geometry of the spatial section: hyperbolic, flat, or closed (spherical), for $A=-1,0$, or 1, respectively. The metric (\ref{ltbm}) is not in the standard spherical form (\ref{metric}), so the general formulae derived in Sec. II do not apply, but it is straightforward to compute $K_{ab}$ directly from its definition. In terms of the $\{x^{\mu}_{-}\}$ coordinates, the shell is given by 
\beq
t=\tau, \;\; r=x_{\rm c}=\mbox{const.}, 
\eeq
and the angular coordinates are, as before, trivially identified. The outward pointing unit normal to $\Sigma$ is
\beq
n^{\mu}_{-}=\fr{\sqrt{1-w_{\rm c}}}{a(\tau)}\delta^{\mu}_{r},
\eeq
where $w_{\rm c}\equiv w(x_{\rm c})=\mbox{const.}$, and continuity of the four-metric across $\Sigma$ implies $R(\tau)=a(\tau)x_{\rm c}$. As before, we only need $K^{\pm}_{\theta\theta}$:
\beq
K_{\theta\theta}^{-}=R\sqrt{1-w_{\rm c}}, \;\;
K_{\theta\theta}^{+}=R\sqrt{1-\fr{2M}{R}+\dot{R}^{2}},
\eeq
where $M$ is the total gravitational mass appearing in the exterior Schwarzschild metric. The equation $[K_{\theta\theta}]=-4\pi\sigma R^{2}$ gives
\beq
\dot{R}^{2}=\tilde{k}^{2}-1+2\fr{M-\mu_{\Sigma}\tilde{k}}{R}+\fr{\mu_{\Sigma}^{2}}{R^{2}}. \label{ed2}
\eeq
where $\mu_{\Sigma}\equiv4\pi\sigma R^{2}$ and $\tilde{k}\equiv\sqrt{1-w_{\rm c}}$. This equation is formally identical to that for the two Schwarzschild shells [cf. Eq. (\ref{dyn2})] provided we make the identifications $\tilde{k}\equiv k$ and $\mu_{\Sigma}\equiv m_{\Sigma}/2=m_{+}/(2k)$. That is, a thin shell of proper mass $m_{\Sigma}$ moving in an interior Schwarzschild background with mass parameter $m_{-}$ is equivalent to a thin shell of proper mass $m_{\Sigma}/2$ moving in a LFRW background with total gravitational mass $m_{-}$. The factor of $1/2$ may be heuristically understood due to the presence of a continuous distribution of matter adjacent to the shell, instead of vacuum: half of the proper mass of the shell in the former case has already been accounted for in the inner side of the shell, in $m_{-}$.

Finally, we note that the usual boundary surface problem of matching a closed LFRW universe to a Schwarzschild exterior~\cite{mtwmatch} is straightforwardly obtained from our construction by setting 
$$
m_{\Sigma}=m_{+}=0\Rightarrow M=m_{-}, \;\; A=1\Rightarrow w_{\rm c}=x_{\rm c}^{2}.
$$
The condition $[K_{ab}]=0$ leads to a single non-trivial equation [cf. Eq. (\ref{ed2})]
\beq
\dot{a}^{2}=2\fr{M}{a}-1,
\eeq
which is just the Friedmann equation for a closed universe ($k=1$) with energy density $\mu=3M/(4\pi x_{\rm c}^{3})$.

\section{Shells with pressure}

We consider here the case of spherical shells with the following surface stress-energy tensor defined on them:
\beq
S_{ab}=(\sigma+p)u_{a}u_{b}+p\gamma_{ab},
\eeq
where $p$ is a tangential pressure, by virtue of spherical symmetry. Physically, this matter configuration can be realized by, e.g., `counter-rotating' dust, where the individual dust particles are taken to follow geodesic orbits on the shell, with half of them orbiting along any given great circle with angular momentum per unit rest mass $l$, and the other half in the opposite direction with angular momentum per unit rest mass $-l$, such that the net angular momentum vanishes, thereby preserving spherical symmetry~\cite{ecluster}.

The junction conditions yield formally the same equation of motion as in the pressureless case [cf. Eq. (\ref{dyn})], but, unlike in the latter case, the proper mass of the shell is no longer conserved. This is the crucial difference, and implies that the ratio $m_{+}/m_{\Sigma}$ is now an implicit function of $R$. The conservation equation (\ref{hamc}) gives
\beq
\dot{\sigma}+2\fr{\dot{R}}{R}(\sigma+p)=0. \label{ce}
\eeq
Note that even the `generalized' mass, $\tilde{m}_{\Sigma}\equiv 4\pi(\sigma+2p)R^{2}$ fails to be conserved in general. If the tangential pressure is constant, then $\tilde{m}_{\Sigma}$ is conserved, and the evolution equation becomes identical to the one for the pressureless case, upon the substitution $m_{\Sigma}\rightarrow \tilde{m}_{\Sigma}$, i.e., the constant pressure case is equivalent to the pressureless one, provided one includes pressure into the generalized mass definition (note, however, that even in this case $\dot{m}_{\Sigma}\neq0$, since there is always work done against pressure `forces').  

The conservation equation (\ref{ce}) can be rewritten as a first-order ODE for $\sigma(R)$, by using $\dot{\sigma}=(d\sigma/dR)\dot{R}$, as
\beq
\fr{d\sigma}{dR}+\fr{2}{R}(\sigma+p)=0. \label{cons}
\eeq
Upon introduction of the dimensionless function $\psi(R)\equiv 2\pi\sigma(R)R$, the equation above can be rewritten as
\beq
\fr{1}{2\pi}\fr{d\psi}{dR}=-(\sigma+2p). \label{psi}
\eeq
Now, one may either (i) specify an equation of state of the form $p=p[\sigma]$ and solve Eq. (\ref{cons}) for $\sigma(R)$, or (ii) prescribe a functional form for $\psi(R)$, thereby fixing $\sigma(R)$, with $p[\sigma]$ following directly from Eq. (\ref{psi}).

The equation of motion for the shell reads then
\beq
\dot{R}^{2}+V(R)=0,
\eeq
with the effective potential
\beq
V(R)=1-\fr{a}{R^{2}\psi^{2}}-\fr{b}{R}-\psi^{2}, \label{potn}
\eeq
where
\beq
a\equiv\left(\fr{m_{+}}{2}\right)^{2},\;\; b\equiv M+m_{-}.
\eeq
The dynamics of the shell will obviously depend on $\psi$ and vice-versa, i.e., a given form of the potential---whereby the dynamics is uniquely determined---will constrain the allowed choices for $\psi$, or equivalently, for the equation of state (although it will not uniquely determine them). The qualitative nature of the evolution is given by the shape of the effective potential: its zeros (if any), and first and second derivatives. From Eqs. (\ref{psi}) and (\ref{potn}), we have then
\bqa
V'&=&2\psi'\left(\fr{a}{R^{2}\psi^{3}}-\psi\right)+\fr{2a}{R^{3}\psi^{2}}+\fr{b}{R^{2}}, \label{vp} \\
V''&=&2\psi''\left(\fr{a}{R^{2}\psi^{3}}-\psi\right)+\fr{4a\psi'}{R^{3}\psi^{3}} \nb \\
&&-\fr{6a}{R^{2}\psi^{2}}\left(\fr{\psi'}{\psi}+\fr{1}{R}\right)^{2}-2\fr{b}{R^{3}}-\psi'^{2}. \label{vp2}
\eqa

Now, in the presence of non-vanishing pressure it is possible to design the shape of $V$, so as to satisfy the necessary and sufficient conditions for a stable solution, $R_{*}$:
\beq
V(R_{*})\leq0,\;\; V'(R_{*})=0,\;\; V''(R_{*})>0.
\eeq
If the first inequality saturates, the solution is stationary. 

\subsection{Non-existence of stable solutions with a strictly linear barotropic equation of state}

{\em Proposition.} There are no stable configurations with a strictly linear barotropic equation of state $p=\alpha\sigma$.

{\em Proof.} Let us assume that $p=\alpha\sigma$, with $\alpha\in(-1,1)$, where the upper and lower limits are imposed by causality. Equation (\ref{psi}) integrates to $\psi=A/R^{1+2\alpha}$, where $A>0$ is an integration constant, and we have then
\bqa
V&=&1-\fr{a}{A^{2}}R^{4\alpha}-\fr{b}{R}-\fr{cA^{2}}{R^{2(1+2\alpha)}}, \\
V'&=&-4\alpha\fr{a}{A^{2}}R^{4\alpha-1}+\fr{b}{R^{2}}+2(1+2\alpha)\fr{A^{2}}{R^{3+4\alpha}}, \\
V''&=&-4\alpha(4\alpha-1)\fr{a}{A^{2}}R^{4\alpha-2}-2\fr{b}{R^{3}} \nb \\
&&-2(1+2\alpha)(3+4\alpha)\fr{A^{2}}{R^{4(\alpha+1)}}. \label{vpp}
\eqa
The condition $V'=0$ leads to a quadratic equation for $X\equiv R^{1+4\alpha}$:
\beq
X^{2}-\fr{bA^{2}}{4a\alpha}X-\fr{(1+2\alpha)A^{4}}{2a\alpha}=0, \label{af}
\eeq
with solution(s):
\beq
X^{\pm}_{0}=\fr{bA^{2}}{8a\alpha}\left(1\pm\sqrt{1+32a(1+2\alpha)\alpha/b^{2}}\right). \label{root}
\eeq
Now, at the local extrema, from Eqs. (\ref{vpp}) and (\ref{af}), after some manipulation we obtain
\beq
V''_{0}=-\fr{4a\alpha(1+4\alpha)}{A^{2}X_{0}^{4(1+\alpha)/(1+4\alpha)}}\left[X^{2}_{0}+\fr{A^{4}(1+2\alpha)}{2a\alpha}\right].
\eeq
The solution $R_{0}\equiv X_{0}^{1/(1+4\alpha)}$ is stable if and only if $V''_{0}>0$, i.e., if and only if either of the following holds:
\bqa
&&\mbox{(i)}\;\; \alpha(1+4\alpha)<0\;\; \mbox{and}\;\; X^{2}_{0}+\fr{A^{4}(1+2\alpha)}{2a\alpha}>0, \nb \\
&&\mbox{(ii)}\;\; \alpha(1+4\alpha)>0\;\; \mbox{and}\;\; X^{2}_{0}+\fr{A^{4}(1+2\alpha)}{2a\alpha}<0. \nb
\eqa
If (i) holds, then $\alpha\in(-1/4,0)$, and we must take $X_{0}^{-}$ in (\ref{root}) with the further condition that the discriminant is greater than unity; however, this requires $\alpha>0$ or $\alpha<-1/2$, both of which contradict $-1/4<\alpha<0$. Now, let us examine possibility (ii): we have either $\alpha>0$ or $\alpha\in(-1,-1/4)$. If $\alpha>0$, the condition $X^{2}_{0}+A^{4}(1+2\alpha)/(2a\alpha)<0$ is violated. If $\alpha\in(-1,-1/2)$, the former condition is also violated, so we are left with $\alpha\in[-1/2,-1/4)$. But this implies (since $\alpha$ is negative) that $X_{0}^{+}<0$, and thus we must take $X_{0}^{-}$ in (\ref{root}) subject to either $\alpha>0$ or $\alpha<-1/2$, both of which are incompatible with $\alpha\in[-1/2,-1/4)$. We conclude, therefore, that there are no solutions with $V'=0$ and $V''>0$. $\square$

It immediately follows that all momentarily static solutions---which exist whenever $V=0$---are unstable, i.e., there are no stationary solutions for shells with a strictly linear barotropic equation of state.

\subsection{Energy conditions and stability criteria}

From Eq. (\ref{vp2}), together with the requirements $V=V'=0$, after some algebra, the necessary and sufficient condition for a stable solution ($V''>0$) may be written as
\bqa
&&-\fr{\psi''}{\psi'}\left(\fr{a}{R^{3}\psi^{2}}+\fr{b}{2R^{2}}\right)>-\fr{2a\psi'}{R^{3}\psi^{3}}+ \nonumber \\
&&\fr{3a}{R^{2}\psi^{2}}\left(\fr{\psi'}{\psi}+\fr{1}{R}\right)^{2}+\fr{b}{R^{3}}+\fr{\psi'^{2}}{2}. \label{stc}
\eqa
Now, from Eqs. (\ref{cons})-(\ref{psi}), we have
\beq
\psi''=-\sigma'\left(1+2\fr{dp}{d\sigma}\right)=\fr{2}{R}(\sigma+p)(1+2C^{2}_{\rm s}), \label{psip}
\eeq
where $C_{\rm s}\equiv \sqrt{dp/d\sigma}$ is the local sound speed, if one regards $p$ as a hydrostatic pressure. The stability condition (\ref{stc}) reads then
\beq
\fr{(\sigma+p)(1+2C_{\rm s}^{2})}{\sigma+2p}>\Theta_{1}\left[\fr{a}{2\pi^{2}\sigma^{2}R^{6}}P(\varsigma)+\Theta_{2}\right], \label{stc2}
\eeq
where $\varsigma\equiv p/\sigma$, $P(\varsigma)=1+2\varsigma+6\varsigma^{2}$, and the $\Theta_{i}$ are strictly positive terms:
\bqa
\Theta_{1}&=&\fr{2\pi^{2}\sigma^{2}R^{5}}{a+2\pi^{2}b\sigma^{2}R^{3}}, \\
\Theta_{2}&=&\fr{b}{R^{3}}+2\pi^{2}(\sigma+2p)^{2}.
\eqa
The quadratic $P(\varsigma)$ has an absolute positive minimum at $\varsigma=-1/6$, and therefore the right-hand-side of (\ref{stc2}) is strictly positive.

Now, in terms of the energy density and principal pressures, the energy conditions for this system are:

\vspace{0.2cm}

{\em Null energy condition (NEC):} $\sigma+p>0$

\vspace{0.2cm}

{\em Dominant energy condition (DEC):} $\sigma>0$, $\sigma>|p|$

\vspace{0.2cm}

{\em Weak energy condition (WEC):} $\sigma>0$, $\sigma+p>0$

\vspace{0.2cm}

{\em Strong energy condition (SEC):} $\sigma+p>0$, $\sigma+2p>0$

\vspace{0.2cm}

The DEC implies the WEC, which in turn implies the NEC, and the SEC also implies the NEC; the energy conditions are otherwise independent. Comparison between these energy conditions and Eqs. (\ref{psi}) and (\ref{psip}), shows that the latter provide a clear relation between the former and the matter content of the shell. Accordingly, one expects the stability, or lack thereof, of the shell to depend crucially on the energy conditions.

For a strictly linear barotropic equation of state, condition (\ref{stc2}) implies $\sigma+p>0$. Clearly, if the WEC is violated, there are no stable configurations. (In fact, as shown in the previous subsection, even if the WEC is satisfied, there are no stable solutions for such equations of state).

For {\em arbitrary} equations of state, one may distinguish two cases, depending on whether $C_{\rm s}^{2}$ is non-negative. Most macroscopic forms of matter obey $C_{\rm s}^{2}\geq0$ [including matter opaque to sound ($C_{\rm s}^{2}=0$), such as tenuous plasmas where electromagnetic radiation below the plasma cut-off frequency does not propagate], whereas certain fundamental fields, such as massive scalar fields, exhibit $C_{\rm s}^{2}<0$. We shall assume throughout that $\sigma>0$, remarking, however, that matter with negative energy density has been considered from a purely theoretical standpoint, e.g., in attempts to construct stable wormhole solutions~\cite{visser95}. We skip trivial algebra and present the relation between energy conditions and stability in Table I below.

\begin{table}[ht] \centering \begin{tabular}{ccccc} 
\hline \hline
$\;\;\;\;C_{\rm s}^{2}\;\;\;\;$  & $\;\;\;$ WEC $\;\;\;$ & $\;\;\;$ SEC $\;\;\;$ & $\;$ Unstable $\;$ \\ 
\hline
$\geq0$ & Yes & No & Yes \\
$\geq0$ & No & No & Possibly \\
$\geq0$ & Yes & Yes & Possibly \\
$<0$ & Yes & No & $-1/2<C_{\rm s}^{2}<0$ \\
$<0$ & No & No & $-1<C_{\rm s}^{2}<-1/2$ \\
$<0$ & Yes & Yes & $-1<C_{\rm s}^{2}<-1/2$ \\
\hline \hline \\
\end{tabular} 
\caption{Stability criteria and energy conditions.}
\end{table}  

\section{Trapped surfaces}

Let $\Sigma$ be any given compact spatial two-surface in the spacetime, and $\Theta_{\pm}$ be the expansions in the future-pointing null directions orthogonal to $\Sigma$. The latter is said to be a trapped surface if $\Theta_{+}\Theta_{-}\geq0$~\cite{penrose73}. Trapped surfaces signal thus the boundary of a region where any initially expanding null congruence begins to converge; clearly, they define regions of `no escape'. The limiting case where expansion vanishes along future-oriented normal null directions is referred to as an {\em outer marginally trapped surface} (OMTS). It is well known that the existence of trapped surfaces implies that of OMTS~\cite{wald84}. In spherical symmetry, the apparent horizon (AH)---the outer boundary of a compact trapped region---is an OMTS.

Let us consider the case of an arbitrary shell $\Sigma$ matched to two arbitrary spherical spacetimes, as described in Sec. II. Take the vacuum metric to be of the form (\ref{metric}), with $g_{rr}=1/F_{\pm}(t,r)$, without imposing {\em a priori} any sign constraints on $F$. Since the four-metric $g_{\mu\nu}^{\pm}$ is continuous across $\Sigma$ and one needs derivatives of null vector fields orthogonal to $\Sigma$, it suffices to have $g_{\mu\nu}^{\pm}$ to compute $\Theta$ along ingoing and outgoing null directions. Our surface $\Sigma$ is that of a spherical shell with proper area radius $r=R(\tau)$ (which provides a coordinate invariant definition, and, in addition, allows for measurements by an external observer, who can measure proper circumferences, but not coordinate radii). It is convenient to introduce null coordinates $\{u,v\}$ defined by
\beq
du=hdt-\fr{1}{\sqrt{F}}dr,\;\; dv=hdt+\fr{1}{\sqrt{F}}dr,
\eeq
from which it follows that
\beq
\Theta_{\pm}=\left(\fr{1}{h}\fr{\partial}{\partial t}\mp\sqrt{F}\fr{\partial}{\partial r}\right)R(\tau),
\eeq
where the `$\pm$' sign denotes evaluation along outgoing and ingoing null directions, respectively. The condition for trapped surfaces reads then [where Eq. (\ref{eta}) was used]
\beq
\Theta_{+}\Theta_{-}=\fr{\dot{R}^{2}}{1+\dot{R}^{2}/F}-F\geq0. \label{ahc}
\eeq
The AH is given by the limit $\Theta_{+}\Theta_{-}\rightarrow0$, which requires $F\rightarrow0$.
For the case of a boundary surface, $[K_{ab}]=0$, the time coordinate is globally defined, and thus $F_{-}=F_{+}$ {\em everywhere}, thereby uniquely defining a continuous curve $t_{\rm ah}(r)$, via $F_{\pm}=0$. In the presence of a thin shell, however, the functional forms of $F_{-}$ and $F_{+}$ are different, and therefore the AH curve (implicitly defined by two {\em different} curves in ${\mathcal M}_{\pm}$, by $F_{\pm}=0$, respectively) may not necessarily be continuous.

\begin{figure}
\begin{center}
\epsfysize=15pc
\epsfxsize=22pc
\epsffile{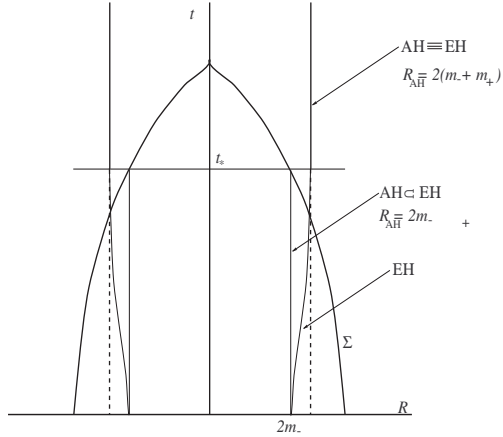}
\end{center}
\caption{Apparent (AH) and event horizon (EH) evolution for a thin shell $\Sigma$ with active gravitational mass $m_{+}$ imploding in an interior Schwarzschild background with mass parameter $m_{-}$.\label{fig4}}
\end{figure}

To illustrate this, let us consider the case of a single shell with active gravitational mass $m_{+}$ moving in an interior Schwarzschild background with mass parameter $m_{-}$. This case is of sufficient generality, since, as shown in the preceding sections, the vacuum-immersed single shell case, as well as that of a shell collapsing onto an interior LTB solution, can be trivially obtained from it. We have then
\beq
F_{-}=1-\fr{2m_{-}}{r}=0, \;\; F_{+}=1-2\fr{m_{+}+m_{-}}{r}=0,
\eeq
which defines the AH as
\beq
t_{\rm ah}(r)=2m_{+}H_{1}(r-R)+2m_{-},
\eeq
where $H_{1}(x)$ is the unit Heaviside step-function ($H_{1}=0$ for $x<0$, and $H_{1}=1$, for $x\geq 0$). The evolution of the AH is shown in Fig. \ref{fig4}.

The AH coincides initially with the EH, since the inner spacetime is just static Schwarzschild, but {\em jumps out discontinuously} at $t=t_{*}$, when the thin matter shell $\Sigma$ crosses the initial Schwarzschild radius. The EH is determined by the entire future development (along null generators) of the spacetime, and is therefore {\em continuous}, unlike the AH, which is defined locally in terms of the gravitational mass interior to a certain radius, on a given spacelike hypersurface.

\section{Shell crossing}

In spherical gravitational collapse, one may distinguish between two different types of singularities, given by the location and relative motion of neighboring shells. A {\em shell focusing} singularity is said to occur at the center of symmetry, when the proper area radius vanishes therein, leading to the blow-up of all curvature invariants; these singularities are gravitationally strong (any volume form defined along the Jacobi fields vanishes at the singularity), and the spacetime is geodesically incomplete~\cite{central}. A {\em shell crossing} singularity is said to occur when neighboring shells cross each other at finite (comoving) radius, thereby leading to two-dimensional caustics, where the energy density diverges, and the metric becomes singular~\cite{yodzis73}. In LTB collapse, shell crossing occurs for spatially inhomogeneous density distributions, whenever the proper time for collapse of a given shell is a monotonically decreasing function of the comoving radius. For the charged dust case, shell crossing has been shown to be inevitable, both for asymptotically flat~\cite{ori91}, and asymptotically de Sitter~\cite{goncalves01a} spacetimes. Physically, shell crossings signal the intersection of matter flow lines at a given spacelike surface, to the future of which the model therefore breaks down. Such shell crossing singularities have been shown to be gravitationally weak~\cite{newman86}, and, in one particular LTB case, an analytical metric continuation was found in a distributional sense~\cite{papapetrou&hamoui67}. Recently, Szekeres and Lun~\cite{szekeres&lun99} showed that it is always possible to find a coordinate transformation that renders the metric $C^{1}$ (but not $C^{2}$) at the LTB shell crossing singularity. Their result adds considerably to our understanding of shell crossing in LTB collapse, but there remains to be shown that such $C^{1}$ transformation always exists (and is presumably unique) for a generic spherical metric. 

\subsection{Conditions for shell crossing}

One might hope that a detailed analysis of the dynamics of individual thin shells could shed some light on the shell crossing process in continuous matter models. Thus motivated, we shall consider the case of two Schwarzschild shells, as described in Sec. III. Take the shells to have initial radii $R_{0}^{\pm}$, and dynamics given by 
\beq
\dot{R}^{\pm}+V^{\pm}(R^{\pm})=0, \label{deq3}
\eeq
where
\bqa
&&V^{\pm}=-\sqrt{k^{2}-1+\fr{a_{\pm}}{R^{\pm}}+\left(\fr{b_{\pm}}{R^{\pm}}\right)^{2}}, \\
&&a_{-}=2k\sqrt{b_{-}}=m_{-}, \nb \\
&&a_{+}=2(k\sqrt{b_{+}}+m_{-})=M+m_{-}. \nb
\eqa
For simplicity, and without loss of generality, we shall consider the marginally bound case, $k=1$. In this case, the dynamical equation (\ref{deq3}) admits an analytical solution $R^{\pm}(\tau)$, which is explicitly given in Appendix A. The proper time for collapse is given by solving $R(\tau_{\rm c})=0$, which yields
\beq
\tau_{\rm c}=\fr{2}{3a^{2}}\left(2b^{3/2}+a^{3/2}\sqrt{\fr{4b^{3}}{a^{3}}-\fr{3bR_{0}^{2}}{a}+R_{0}^{3}}\right), \label{tcol}
\eeq
where the `$\pm$' subscripts have been omitted for clarity. For given gravitational masses $m_{\pm}$ (whereby the corresponding $a_{\pm},b_{\pm}$ are determined), the proper time for collapse is a function of the initial radius $R^{\pm}_{0}$. Now, shell crossing occurs iff, for $R_{0}^{+}>R_{0}^{-}$, we have 
\beq
\tau_{\rm c}(R^{+}_{0})<\tau_{\rm c}(R^{-}_{0}). \label{sxco}
\eeq
For given $m_{\pm}$, it is straightforward to numerically solve for condition (\ref{sxco}), thereby obtaining a region of the two-dimensional parameter space---given by the half-plane $\{R_{0}^{-}\in(0,+\infty)\}\times\{R_{0}^{+}>R_{0}^{-}\}$---in which (\ref{sxco}) is satisfied. One can actually obtain an analytical criterion for shell crossing, by restricting our attention to {\em neighboring} shells, defined by
\beq
R_{0}^{+}=R_{0}^{-}(1+\xi),
\eeq
with $0<\xi\ll1$. For shells thus defined, we can Taylor expand 
\beq
\tau_{\rm c}(R_{0}^{+})=\tau_{\rm c}(R_{0}^{-})+\left(\fr{d\tau_{\rm c}}{dR_{0}}\right)_{R_{0}^{-}}\xi+{\mathcal O}(\xi^{2}).
\eeq
Condition (\ref{sxco}) reads then
\beq
\left(\fr{d\tau_{\rm c}}{dR_{0}}\right)_{R_{0}^{-}}<0.
\eeq
From Eq. (\ref{tcol}), after a little algebra, this gives
\beq
R^{-}_{0}\in(0,R_{*}),
\eeq
where $R_{*}\equiv2b_{-}/a_{-}=m_{-}/2$. We note that this (necessary and sufficient) condition for shell crossing for neighboring shells requires the inner shell to be inside its Schwarzschild radius. From the definition of neighboring shells, it follows that the outer shell is also inside its Schwarschild radius: Take the limiting case $R_{0}^{-}=m_{-}/2$, and set $R_{0}^{+}=m_{-}/2+\delta$, with $\delta\sim{\mathcal O}(m_{-}^{-1})$. Suppose that $R_{0}^{+}$ is outside its Schwarzschild radius, i.e., $R_{0}^{+}>R_{\rm Sch}^{+}=2(m_{+}+m_{-})$. This implies $$4m_{+}+3m_{-}<2\delta,$$
which cannot be, since $m_{+}>0$ and $\delta\sim {\mathcal O}(m_{-}^{-1})$; hence, $R_{0}^{+}<R^{+}_{\rm Sch}$. Note that the converse is not necessarily true: one may have $R_{0}^{+}>R_{\rm Sch}^{+}$ and $R_{0}^{-}<R_{\rm Sch}^{-}$, provided $m_{+}\lesssim{\mathcal O}(m_{-}^{-1})$.

\begin{figure}
\begin{center}
\epsfysize=11pc
\epsfxsize=15pc
\epsffile{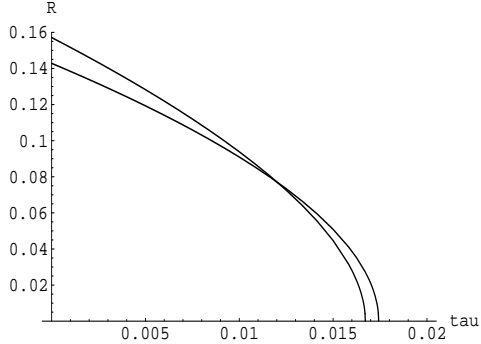}
\end{center}
\caption{Gravitational collapse for two neighboring shells with data $m_{+}=m_{-}=1$, $R^{-}_{0}=1/7$, $R^{+}_{0}=R^{-}_{0}(1+10^{-1})$. The condition $R_{0}^{-}<R_{0}^{+}<R_{*}$ (where $R_{*}=1/6$) is obeyed, and hence the two shells cross before collapsing to zero radius. Note that {\em both} shells are initially inside the respective Schwarzschild radii. \label{fig5}}
\end{figure}

The case of non-neighboring shells [i.e., $R_{0}^{+}-R_{0}^{-}\sim {\mathcal O}(R_{0}^{-})$] is qualitatively analogous: shell crossing only occurs if either (i) both shells are ``sufficiently'' inside their respective Schwarzschild radii, or (ii) the inner shell is initially untrapped, and the outer shell ``sufficiently'' trapped~\footnote{The upper limit for the initial radius of the outer shell depends on the initial radius for the inner shell, and on both masses $m_{\pm}$.}. Illustrative examples are shown below.

\begin{figure}
\begin{center}
\epsfysize=11pc
\epsfxsize=15pc
\epsffile{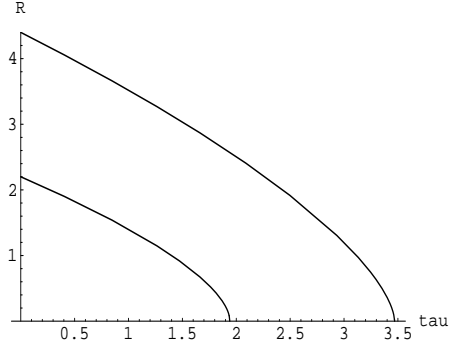}
\end{center}
\caption{Gravitational collapse for two shells with data $m_{+}=m_{-}=1$, $R^{-}_{0}/R^{-}_{\rm Sch}=R^{+}_{0}/R^{+}_{\rm Sch}=1.1$. Both shells are initially {\em outside} their Schwarzschild radii, and, as such, fail to obey the shell crossing.\label{fig6}}
\end{figure}

\begin{figure}
\begin{center}
\epsfysize=11pc
\epsfxsize=15pc
\epsffile{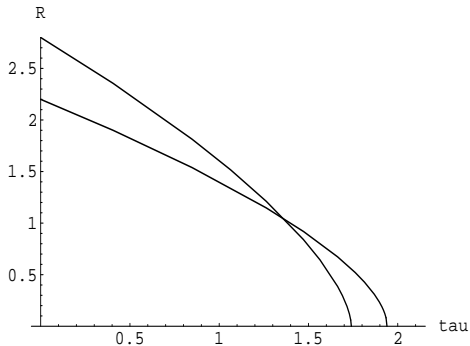}
\end{center}
\caption{Gravitational collapse for two shells with data $m_{+}=m_{-}=1$, $R^{-}_{0}/R^{-}_{\rm Sch}=1.1$ and $R^{+}_{0}/R^{+}_{\rm Sch}=0.7$. The outer shell is initially inside its Schwarzschild radius, but not the inner one. In this case, the shells cross at finite radius. \label{fig7}}
\end{figure}

\subsection{Discrete shell model vs. LTB collapse}

In a multi-thin-shell model, the metric is at least $C^{0}$ at the shell crossing by construction, since $[g_{\mu\nu}]=0$ for all times. This is in {\em apparent} contradiction with the well-known result for LTB collapse, where the comoving metric becomes singular ($g_{rr}\rightarrow0$) at shell crossings~\cite{newman86}. Whilst such a metric singularity may be removed by a $C^{1}$ coordinate transformation via the prescription of Szekeres and Lun~\cite{szekeres&lun99}, one is still left with the fact that scalar quantities such as the energy density remain divergent at LTB shell crossings (in fact, tidal forces are still infinite thereon), whereas they are finite in the discrete shell case [in particular, $\sigma=m/(4\pi R^{2}_{\rm sc})<\infty$]. Accordingly, the former cannot be taken as the continuous limit of the latter.

The inequivalence---as far as shell crossings are concerned---between these two models also arises from the following observation: In LTB collapse, individual shells follow geodesic motion in the four-dimensional spacetime due to the lack of any external forces (collapse is pressureless and uncharged), but this does not happen in the present case; within each shell, particle motion is geodesic, but, from the viewpoint of the four-dimensional spacetime, the shell is accelerated:
\bqa
a^{\mu}&=&u^{\nu}\nabla_{\nu}u^{\mu}=\dot{R}\left[\fr{\ddot{R}}{\eta}+\fr{h'}{h}\left(2\eta-\fr{1}{\eta h^{2}}\right)\right]\delta^{\mu}_{t} \nb \\
&&+\left(\ddot{R}+\fr{h'}{h^{3}}\right)\delta^{\mu}_{r}\neq 0, \\
a^{a}&=&u^{b}\,^{(3)}\nabla_{b}u^{a}=\delta^{b}_{\tau}\,^{(3)}\nabla_{b}\delta^{a}_{\tau}=0.
\eqa

We remark that the shell crossing process {\em cannot} be determined from the field equations alone, as a matter of principle. Once two (or more) shells cross, one must specify {\em a priori} the type of interaction that takes place (which depends on the microphysics of the model at hand), in order to follow the dynamics into the causal future of the shell cross surface. Only in special cases, where a {\em purely gravitational gravitational interaction} (whence the rest mass of each shell is conserved) between different shells is assumed, can one determine the evolution of the system beyond the shell cross. For null surfaces, we have the well-known DTR (Dray-t'Hooft-Redmount) relations~\cite{dtr}, which relate the mass and momenta of the different regions before and after shell crossing. Such relations were subsequently generalized to the timelike case (of massive spherical shells) by N\'{u}\~{n}ez, Oliveira, and Salim~\cite{nunez&oliveira&salim93}.

\section{Concluding remarks}

We have analyzed in some detail the dynamics, stability, trapped surfaces, and shell crossing for spherical thin shells in vacuum. By focusing on the dynamics of the exterior shell in a vacuum two-shell model, the problem reduces to that of a single shell moving freely (albeit not geodesically) in a Schwarzschild background with different active mass parameters on each side.

We have shown that there can be no stable solutions with a strictly linear barotropic equation of state, regardless of any energy conditions satisfied by such matter fields. Whereas this rules out several idealized forms of matter as candidates for stable shell configurations, it still leaves ample room for stability with more general equations of state. For instance, by considering a quasilinear equation of state (which is a good approximation to any equation of state near equilibrium, if the latter exists), Brady, Louko, and Poison~\cite{brady&louko&poison91} showed that stable {\em static} shells with an interior Schwarzschild background can exist, provided certain conditions are satisfied, which turn out to supersede energy conditions necessary for the existence (but not necessarily stability) of such shells~\cite{frauendiener&hoenselaers&konrad90}. For matter configurations with $C_{\rm s}^{2}\geq0$ obeying the WEC, but otherwise arbitrary, we showed here that the SEC is a {\em necessary and sufficient} condition for the existence of stable shells around a classical Schwarzschild black hole, thereby producing a very simple and useful test for the stability of {\em any} matter content. For matter with $C_{\rm s}^{2}\in(-1/2,0)$ obeying the WEC, the SEC is also necessary and sufficient for stability. For $C_{\rm s}^{2}\in(-1,-1/2)$ the analysis cannot be carried out further without specifying an equation of state, to evaluate inequality (\ref{stc2}). However, the results of~\cite{brady&louko&poison91} for static shells with $C_{\rm s}^{2}<0$ suggest that there are no stable configurations, even dropping the staticity assumption.

The collapse of a thin matter shell onto a Schwarzschild black hole introduces a step-function-type discontinuity in the apparent horizon curve, which occurs when the shell crosses the initial Schwarzschild radius, whence the jump equals the (active) gravitational mass of the shell. 

Neighboring dust shells were shown to cross whenever the inner shell is sufficiently inside its Schwarzschild radius, specifically, $R_{0}^{-}<R^{-}_{\rm Sch}/4$. Shell crossings occur in a multi-dust-shell case, just as they do in the continuous LTB dust case, but individual shells move geodesically in the latter, whereas they are accelerated in the former. This, together with the fact that the energy density remains finite in the discrete case, but diverges in the LTB one, implies that the multi-shell case cannot be taken as the discrete analogue of LTB collapse, insofar as shell crossings are concerned.

\begin{acknowledgments}
It is a pleasure to thank Vince Moncrief for several helpful discussions. This work was supported by FCT Grant SFRH-BPD-5615-2001 and by NSF Grant PHY-0098084.
\end{acknowledgments}

\appendix

\section{Marginally bound collapse of two dust shells}

Formally, we need to solve the equation
\beq
\dot{R}+\sqrt{\fr{a}{R}+\fr{b}{R^{2}}}=0,
\eeq
with the initial condition $R(0)=R_{0}$.
The positive real solution is
\beq
R(\tau)=\fr{b}{a}-2\fr{b^{2}}{a}\chi^{-1/3}(\tau)-\fr{1}{2a}\chi^{1/3}(\tau),
\eeq
where
\bqa
\chi&\equiv&8b^{3}-4a^{3}\Sigma^{2}+12a^{7/2}\Sigma\tau-9a^{4}\tau^{2}+3a^{2}(\tau-\fr{2\Sigma}{3\sqrt{a}}) \nb \\
&&\times(-16b^{3}+4a^{3}\Sigma^{2}-12a^{7/2}\Sigma\tau+9a^{4}\tau^{2})^{1/2}, \\
\Sigma&\equiv&\left(4\fr{b^{3}}{a^{3}}-3\fr{b}{a}R_{0}^{2}+R_{0}^{3}\right)^{1/2}.
\eqa

When $b=0$ (i.e., $m_{-}=0$), corresponding to a single shell $m_{+}$ in vacuum, the above solution reduces to the well-known marginally bound case of LTB collapse:
\beq
R(\tau)=\left(\fr{3}{2}\right)^{2/3}\left(\fr{4R_{0}^{3}}{9}-\fr{4}{3}\sqrt{a}R_{0}^{3/2}\tau+a\tau^{2}\right)^{1/3}.
\eeq

\end{document}